ORIGINAL RESEARCH PAPER

# Dilution of precision for time difference of arrival with station deployment

Fengyun Zhang[1] | Hao Li[2] | Yulong Ding[1,3,4] | Shuang-Hua Yang[1] | Li Yang[5]

[1]Department of Computer Science and Engineering, Southern University of Science and Technology, Shenzhen, Guangdong, China

[2]Science and Technology on Near-surface Detection Laboratory, Wuxi, Jiangsu, China

[3]Academy for Advanced Interdisciplinary Studies, Southern University of Science and Technology, Shenzhen, Guangdong, China

[4]Department of Mathematics, Southern University of Science and Technology, Shenzhen, Guangdong, China

[5]Army Engineering University of PLA, Nanjing, Jiangsu, China

**Correspondence**

Shuang-Hua Yang, Department of Computer Science and Engineering, Southern University of Science and Technology, Shenzhen, Guangdong, China.
Email: yangsh@sustech.edu.cn

Li Yang, Army Engineering University of PLA, Nanjing, Jiangsu, China.
Email: yangli.aeup@gmail.com

**Funding information**

Scientific Research Foundation of Science and Technology on Near-Surface Detection Laboratory of China, Grant/Award Number: TCGZ2018A006; National Natural Science Foundation of China, Grant/Award Numbers: 61873119, 61911530247; Science and Technology Innovation Commission of Shenzhen, Grant/Award Number: KQJSCX20180322151418232

**Abstract**

A study is conducted with the aim to reveal the relationship between the performance of moving object tracking algorithms and tracking anchor (station) deployment. The dilution of precision (DoP) for the time difference of arrival (TDoA) technique with respect to anchor deployment is studied. Linear and non-linear estimators are used for TDoA algorithms. The research findings for the linear estimator indicate that the DoP attains a lower value when other anchors are scattered around a central anchor; for the non-linear estimator, the DoP is optimal when the anchors are scattered around the target tag. Experiments on both algorithms are conducted that target location precision related to anchor deployment in practical situations for tracking moving objects integrated with a Kalman filter in an ultra-wideband (UWB)-based real-time localization system. The work provides a guideline for deploying anchors in UWB-based tracking systems.

## 1 | INTRODUCTION

Time difference of arrival (TDoA) is a real-time location technique. The typical system setting is a set of known base stations (anchors) and the target to be located. The target transmits signals to (or receives them from) the station at a certain rate. The TDoA technique can derive the target location without time synchronization between the target and the base stations by differencing the receiving time-stamp of each pair of stations to offset the absolute time base.

The algorithms for TDoA techniques are generally more complex than other techniques such as time of arrival, angle of arrival, and received signal strength [1, 2]. In [3–6], several linear estimators are proposed using the linearization technique, which provides a closed-form solution to the problem.

The most commonly used method is the iterative algorithm based on non-linear least squares (NLSs) and its variations, such as Taylor series estimation in [7] and (weighted) Gauss–Newton in [8]. The TDoA technique has a long history and has been adopted in several systems since the last century: the LORAN system [3], acoustic navigation systems [9], GPS etc. From 2000 onward, there has been an emerging trend of adopting the TDoA technique into in-door location systems [10], especially with ultrawideband (UWB) signals [11, 12]. The benefit of this synchronization-free technique and low latency signal in-door location system with UWB signal and TDoA technique is the attainment of low latency and high precision in real system deployment [13–17]. In navigation systems like GPS, the (geometric) dilution of precision (DoP) is a measurement of location error compared with sensor data error. We define the DoP for in-door location systems as follows:

**Definition 1**

$$Location\ Error = DoP \cdot Measurement\ Error. \quad (1)$$

We choose two algorithms and adopt typical linear and non-linear estimators and analyze their DoPs related to the deployment of the anchor location. It turns out that the resulting DoPs have quite different properties. The linear estimator requires an anchor within the centre of the other anchors to attain a smaller DoP near that central anchor. On the other hand, the non-linear estimator can attain optimal DoP when the anchors are uniformly distributed from the target.

The DoP for the TDoA technique was studied in [18] for navigation systems. Compared with former systems such as GPS, in-door location systems are more flexible in system deployment. That makes it possible to study DoP with respect to the deployment of the base station [19–21]. Recent research has focussed on system deployment on the complex spatial area [22, 23]. There are also studies simulating the optimal DoPs for different anchor layout shapes [24–26]. However, DoPs for TDoA problems of different estimators and layout shapes remain for future study.

The contributions of this work focus on

(1) generalizing the linear and non-linear TDoA-based tracking algorithms;
(2) defining a mathematic indicator of TDoA tracking precision—DoP; and
(3) revealing the theoretical relationship between the performance of moving object tracking algorithms and tracking anchor (station) deployment.

The rest of the paper is organized as follows. Section 2 introduces the model for TDoA measurement and proposes two location-estimation algorithms. Section 3 analyzes the DoP for each algorithm and performs numerical simulations of the analytical results. Section 4 illustrates experimental results for TDoA systems in tracking moving objects integrated with Kalman filters (KFs) based on various system deployments. Section 5 concludes.

## 2 | PROBLEM FORMULATION

### 2.1 | TDoA model

The mathematical model of TDoA is illustrated in Figure 1. The central point is a tag that could send the positioning frame to the nearby anchors at a certain frequency and maintain the line of sight (LoS) condition.

In two dimensions, assume that $p = (x,y)^T \in R^2$ are the coordinates of the unknown target, and $p_i = (x_i, y_i)^T \in R^2, i = 1, 2, …, n$ are the known coordinates of anchors. Then, in TDoA measurements, we could obtain the time-stamp $\tau_i$ when the anchor receives the signal. In the LoS condition, we could further assume that the measurement of $\widehat{\tau}_i$ satisfies $\widehat{\tau}_i \sim N(\tau_i, \delta_i^2)$. We set $d_i = \|p - p_i\|$ as the distances of anchors from the target. The core formula of TDoA is

$$\widehat{d}_{ij} \triangleq \widehat{d}_i - \widehat{d}_j = c(\widehat{\tau}_i - \widehat{\tau}_j) := c\widehat{\tau}_{ij}, \quad \forall i,j = 1, 2, …, n, \quad (2)$$

where $c$ is the velocity of the electromagnetic wave. In this way, we represent range differences by the TDoA measurements. Unlike the usual range difference location algorithm, the distance $\widehat{d}_{ij}$ here satisfies $\sum_{i,j}\widehat{d}_{ij} \equiv 0$.

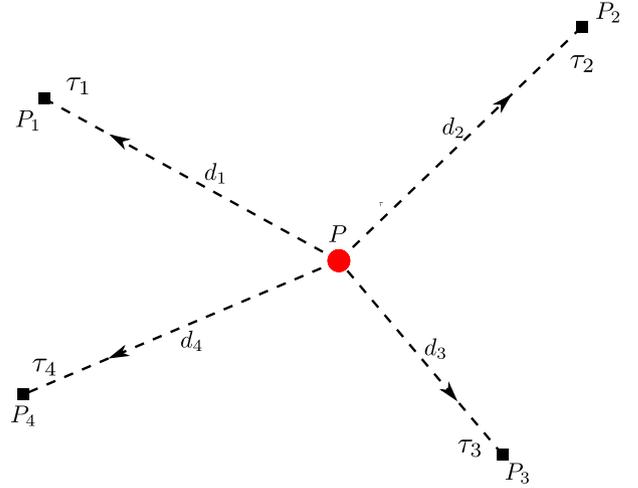

**FIGURE 1** Mathematical model of time difference of arrival

TDoA algorithms aim to find $p$ such that

$$d_{ij}(p) = \widehat{d}_{ij}, \quad \forall\ i,j = 1, 2, …, n, \quad (3)$$

where $d_{ij}(p) = d_i(p) - d_j(p) = \|p - p_i\| - \|p - p_j\|$.

From the foregoing, the inputs of TDoA positioning algorithms are the anchor coordinates $p_i$ and the measured TDoA $\widehat{\tau}_{ij}$, and the output $p$ is the tag's coordinates.

In practice, due to measurement error, the equal sign is usually not satisfied in the previous formula. So the least squares condition is usually considered to estimate position $\widehat{p}$:

$$\widehat{p} = \underset{p}{\operatorname{argmin}} \sum_{i,j \in 1,…,n} \left\| d_{i,j} - \widehat{d}_{i,j} \right\|^2. \quad (4)$$

It is easy to see that the problem belongs to the non-linear least squares model. Furthermore, we can prove that this is a non-convex optimization problem. It is challenging to solve this problem. Thus, iterative algorithms could be adopted, and its convergence and convergence speed should be considered.

Otherwise, since the number of anchors is usually redundant, we could relax some non-linear terms from Equation (3) and obtain a system of linear equations that is overdetermined.

This idea leads to the linear estimator. The system could be solved using the linear least squares method.

## 2.2 | Linear estimator

The core idea of the linear estimator is linearization by relaxing the non-linear term. After that, we could obtain a linear system.

Recall that for any two anchors $i, k$, we have

$$d_{ik} = d_i - d_k = \widehat{d}_{ik}. \quad (5)$$

After squaring, we obtain

$$d_i^2 = \widehat{d}_{ik}^2 + d_k^2 + 2\widehat{d}_{ik}d_k. \quad (6)$$

We can rewrite $d_i^2$ and $d_k^2$ in Equation (6) as follows:

$$d_i^2 = (x_i - x)^2 + (y_i - y)^2$$
$$d_k^2 = (x_k - x)^2 + (y_k - y)^2,$$

In the result, most non-linear terms will be cancelled out with only the non-linear term $d_k$ remaining. Let $r_i^2 = x_i^2 + y_i^2$, $\forall i = 1, 2, \ldots, n$, and Equation (6) becomes

$$r_i^2 - r_k^2 - 2(x_i - x_k)x - 2(y_i - y_k)y = \widehat{d}_{ik}^2 + 2\widehat{d}_{ik}d_k. \quad (7)$$

Note that the non-linear term $d_k$ is the only trouble we face, so we relax this term, that is, treat $d_k$ as another unknown variable. Similarly, for anchors $j$ and $k$,

$$r_j^2 - r_k^2 - 2(x_j - x_k)x - 2(y_j - y_k)y = \widehat{d}_{jk}^2 + 2\widehat{d}_{jk}d_k. \quad (8)$$

Let $(7) \times \widehat{d}_{jk} - (8) \times \widehat{d}_{ik}$, and cancelling the relaxed variable $d_k$, we obtain

$$\begin{aligned}&\widehat{d}_{jk}(r_i^2 - r_k^2) - \widehat{d}_{ik}(r_j^2 - r_k^2) \\ &- 2[\widehat{d}_{jk}(x_i - x_k) - \widehat{d}_{ik}(x_j - x_k)]x \\ &- 2[\widehat{d}_{jk}(y_i - y_k) - \widehat{d}_{ik}(y_j - y_k)]y \\ &= \widehat{d}_{ik}^2 \widehat{d}_{jk} - \widehat{d}_{jk}^2 \widehat{d}_{ik}.\end{aligned} \quad (9)$$

In Equation (9), the anchor $k$ is called the **central anchor**. Without loss of generality, let the central anchor be at the origin $(0, 0)$ and let $k = 0$. Also, note that $\widehat{d}_{jk} - \widehat{d}_{ik} = \widehat{d}_{ji}$, and we have

$$\begin{aligned}&2(\widehat{d}_{j0}x_i - \widehat{d}_{i0}x_j)x + 2(\widehat{d}_{j0}y_i - \widehat{d}_{i0}y_j)y \\ &= \widehat{d}_{j0}r_i^2 - \widehat{d}_{i0}r_j^2 + \widehat{d}_{i0}\widehat{d}_{j0}\widehat{d}_{ji}.\end{aligned} \quad (10)$$

Equation (10) is a linear equation about $(x, y)$, so we could observe that

$$\begin{aligned}A_{ij} &= 2(\widehat{d}_{j0}x_i - \widehat{d}_{i0}x_j) \\ B_{ij} &= 2(\widehat{d}_{j0}y_i - \widehat{d}_{i0}y_j) \\ D_{ij} &= \widehat{d}_{j0}r_i^2 - \widehat{d}_{i0}r_j^2 + \widehat{d}_{i0}\widehat{d}_{j0}\widehat{d}_{ji}.\end{aligned} \quad (11)$$

We could then write Equation (10) as

$$A_{ij}x + B_{ij}y = D_{ij}. \quad (12)$$

For any two anchors $i, j$, we have the formula above, so in the system with $n + 1$ anchors, we could obtain $C_n^2$ linear equations as a linear system:

$$Mp = f, \quad (13)$$

where

$$\begin{aligned}M &= \begin{bmatrix} A_{ij} & B_{ij} \end{bmatrix}_{1 \leq i < j \leq n} \\ f &= \begin{bmatrix} D_{ij} \end{bmatrix}_{1 \leq i < j \leq n}.\end{aligned}$$

In general, the above linear systems are overdetermined. Therefore, when $M^T M$ is invertible, we can use least squares to estimate $p$:

$$\widehat{p} = (M^T M)^{-1} M^T f.$$

The details of the algorithm are shown in Algorithm 1.

---

**Algorithm 1** Algorithm with the linear estimator (with the central anchor)

**Require:** Coordinate $(x_i, y_i)$ for each anchor and measured distance differences $\widehat{d}_{ij}$ between the tag and each pair of anchors
**Ensure:** The fitted position $(x, y)$ of the tag
$M \leftarrow \begin{bmatrix} 2(\widehat{d}_{j0}x_i - \widehat{d}_{i0}x_j) & 2(\widehat{d}_{j0}y_i - \widehat{d}_{i0}y_j) \end{bmatrix}_{1 \leq i < j \leq n}$
$f \leftarrow \begin{bmatrix} \widehat{d}_{j0}r_i^2 - \widehat{d}_{i0}r_j^2 \widehat{d}_{i0}^2 \widehat{d}_{j0} \widehat{d}_{ji} \end{bmatrix}_{1 \leq i < j \leq n}$
$\begin{bmatrix} x \\ Y \end{bmatrix} \leftarrow (M^T M)^{-1} M^T f$

---

### 2.2.1 | Avoiding selection of central anchor

Algorithm 1 will encounter the problem of selecting the central anchor. We shall show that this can be avoided. First, we note that the identity $\widehat{d}_{ik} - \widehat{d}_{jk} = \widehat{d}_{ij}$ and $\widehat{d}_{ik} = -\widehat{d}_{ki}$. Then in Equation (9), the coefficient of $x$ becomes

$$\begin{aligned}&-2[\widehat{d}_{jk}(x_i - x_k) - \widehat{d}_{ik}(x_j - x_k)]\\ &= -2[\widehat{d}_{jk}x_i - \widehat{d}_{ik}x_j + (\widehat{d}_{ik} - \widehat{d}_{jk})x_k] =\\ &= -2(\widehat{d}_{jk}x_i + \widehat{d}_{ki}x_j + \widehat{d}_{ij}x_k)x,\end{aligned}$$

and a similar argument applies to the coefficient of $y$. For the right-hand side, we have

$$\begin{aligned}&\widehat{d}_{ik}^2\widehat{d}_{jk} - \widehat{d}_{jk}^2\widehat{d}_{ik} - (\widehat{d}_{jk}(r_i^2 - r_k^2) - \widehat{d}_{ik}(r_j^2 - r_k^2))\\ &= (\widehat{d}_{ik} - \widehat{d}_{jk})\widehat{d}_{ik}\widehat{d}_{jk} - [\widehat{d}_{jk}r_i^2 - \widehat{d}_{ik}r_j^2 + (\widehat{d}_{ik} - \widehat{d}_{jk})r_k^2]\\ &= -\widehat{d}_{ij}\widehat{d}_{jk}\widehat{d}_{ki} - (\widehat{d}_{jk}r_i^2 + \widehat{d}_{ki}r_j^2 + \widehat{d}_{ij}r_k^2).\end{aligned}$$

So Equation (9) could be changed to

$$\begin{aligned}&2(\widehat{d}_{jk}x_i + \widehat{d}_{ki}x_j + \widehat{d}_{ij}x_k)x\\ &+ 2(\widehat{d}_{jk}y_i + \widehat{d}_{ki}y_j + \widehat{d}_{ij}y_k)y\\ &= \widehat{d}_{jk}r_i^2 + \widehat{d}_{ki}r_j^2 + \widehat{d}_{ij}r_k^2 + \widehat{d}_{ij}\widehat{d}_{jk}\widehat{d}_{ki},\end{aligned}$$

which is again a linear system of $x$ and $y$. The details of the algorithm are shown in Algorithm 2.

---

**Algorithm 2** Algorithm with the linear estimator (without the central anchor)

---

**Require:** Coordinate ($x_i$, $y_i$) for each anchor and measured distance differences $\widehat{d}_{ij}$ between the tag and each pair of anchors.
**Ensure:** The fitted position ($x$, $y$) of the tag.
1: $M \leftarrow [2(\widehat{d}_{jk}x_i\widehat{d}_{ki}x_j\widehat{d}_{ij}x_k)$
    $2(\widehat{d}_{jk}y_i\widehat{d}_{ki}y_j\widehat{d}_{ij}y_k)]_{\text{combination of } i,j,k}$
2: $f \leftarrow [\widehat{d}_{jk}r_i^2\widehat{d}_{ki}r_j^2\widehat{d}_{ij}r_k^2\widehat{d}_{ij}\widehat{d}_{jk}\widehat{d}_{ki}]_{\text{combination of } i,j,k}$
3: $\begin{bmatrix}x\\Y\end{bmatrix} \leftarrow (M^TM)^{-1}M^Tf$

---

Algorithms 1 and 2 only involve matrix multiplication and the inverse and thus are computationally friendly and suitable for the embedded system. The size of $M$ in Algorithm 1 increases linearly with respect to the number of anchors used ($O(n)$) and increases in factorial ($O(n!)$) for Algorithm 2. This suggests not using too many anchors at a time when adopting Algorithm 2.

## 2.3 | Non-linear estimator

The non-linear estimator follows from Equation (3), in which we aim to find $p$ to make Equation (3) hold.

The key idea of solving the non-linear estimator is to use the NLSs model to solve the following optimization problems:

$$\underset{p}{\arg\min} \sum_{1 \leq i < j \leq n} \left\|d_{i,j}(p) - \widehat{d}_{i,j}\right\|^2.$$

Common optimization algorithms include Newton, Taylor expansion, Gauss–Newton, and Levenberg–Marquardt [27]. Compared with other algorithms, Gauss–Newton has the advantage of computing speed without involving the second derivative.

### 2.3.1 | Gauss–Newton iteration

The architecture of Gauss–Newton could be described as follows: for any second derivative bounded vector function $f(x) : \mathbb{R}^m \to \mathbb{R}^n$, and we need to find an optimal $x^*$, such that $\|f(x^*) - y\|$ is minimal. Let $f(x) - y := r(x)$ be the residual item. We assume the modulus of the residual term is close to 0 near the solution. The problem could be rewritten as

$$\arg\min_x \frac{1}{2}\|r(x)\|^2.$$

Let $F(x) = \frac{1}{2}\|r(x)\|^2$. Since the first derivation of $F(x)$ is continuous, $F(x)$ satisfies the condition $\nabla F(x^*) = 0$ at extreme point $x^*$. So the problem can be transformed into zero-point finding problems for $\nabla F(x)$.

If the Newton method is adopted to solve this problem, then given any initial point $x_0$, we have the iterative formula

$$x_{k+1} = x_k - (\nabla^2 F(x_k))^{-1} \nabla F(x_k) \qquad (14)$$

Given $F(x) = \frac{1}{2}\|r(x)\|^2$, we have

$$(\nabla F)_j = \sum_{i=1}^n \frac{df_i}{dx_j} r_i, \qquad (15)$$

so $\nabla F = \mathbf{J}^T r$, where $\mathbf{J}_{ij} = \frac{df_i}{dx_j}$ is the Jacobi matrix of $f$. Next,

$$\begin{aligned}(\nabla^2 F)_{jk} &= \frac{d(\nabla F)_j}{dx_k}\\ &= \frac{d\left(\sum_{i=1}^n \frac{df_i}{dx_j} r_i\right)}{dx_k}\\ &= \sum_{i=1}^n \frac{df_i}{dx_j}\frac{df_i}{dx_k} + \sum_{i=1}^n \frac{d^2f_i}{dx_jdx_k} r_i\end{aligned}$$

so we could write

$$\nabla^2 F = \mathbf{J}^T\mathbf{J} + \sum_{i=1}^{n}\mathbf{H}_i r_i \quad (16)$$

where $\mathbf{H}_i$ is the Hessian matrix of $f_i$.

Because we assume that the residual term near the solution is close to 0, $\mathbf{H}_i r_i$ of Equation (16) could be ignored. If we then put our new Equation (15): $\nabla F(x_k)$ and Equation (16): $\nabla^2 F(x_k)$ into Newton iterative Equation (14), we obtain the Gauss–Newton iterative formula:

$$x_{k+1} = x_k - (\mathbf{J}^T\mathbf{J})^{-1}\mathbf{J}^T r(x_k). \quad (17)$$

### 2.3.2 | TDoA algorithm by Gauss–Newton

When the Gauss–Newton method is adopted to solve the problem for $f(p) = [d_{ij}]_{1\leq i<j\leq n}$, $\mathbf{J}$ becomes

$$\mathbf{J} = \left[\frac{x-x_i}{d_i} - \frac{x-x_j}{d_j} \quad \frac{y-y_i}{d_i} - \frac{y-y_j}{d_j}\right]_{1\leq i,j\leq n} \quad (18)$$

We can now substitute Equation (18) into Equation (17) to obtain the iterative formula. If $\mathbf{J}$ is full rank, the solution of the problem can be iterated through this formula starting from the initial value $p^{(0)}$. The details are shown in Algorithm 3.

---

**Algorithm 3 Gauss–Newton algorithm for the nonlinear estimator**

```
Require: Coordinate (xᵢ, yᵢ) for each anchor
  and measured distance differences d̂ᵢⱼ
  between the tag and each pair of anchors
Ensure: The fitted position (x, y) of the tag
 1: Initialize search point x, y
 2: Repeat:
 3: dᵢ ← √((x − xᵢ)²(y − yᵢ)²)   1 ≤ i ≤ n
 4: r ← [(dᵢ − dⱼ) − d̂ᵢⱼ]_{1≤i,j≤n}
 5: J ← [ (x−xᵢ)/dᵢ − (x−xⱼ)/dⱼ    (y−yᵢ)/dᵢ − (y−yⱼ)/dⱼ ]_{1≤i,j≤n}
 6: [Δx; Δy] ← −(JᵀJ)⁻¹Jᵀr
 7: [x; y] ← [x; y][Δx; Δy]
 8: Until: Δx and Δy are smaller than error
  tolerance
```

---

The Gauss–Newton method will guarantee convergence if $\mathbf{J}$ is well conditioned [28] and the convergence rate is at least linear. In practice, when this estimator is integrated with KF or extended Kalman filter (EKF), only one 'iteration' is needed per step. Thus, the complexity is directly associated with the size of $M$, which grows in factorial with respect to the number of anchors.

## 3 | ANALYSIS OF DoP FOR TDoA ALGORITHM

In this section, we will conduct theoretical analyses on the two TDoA algorithms proposed in Section 2 and give the relationship between the algorithm's solution error and stability as well as the location of the anchors.

### 3.1 | Preliminary

First, we must make some necessary transformations to the whole system for easy analysis. From a physical point of view, the TDoA system should have translation invariance, that is, if the coordinate $p_i$ of all anchors is carried out in some translation transformation, the coordinate $p$ of the solved tag should also be carried out in the same transformation. As a TDoA technique is adopted for positioning, most quantities are formed by the difference of coordinates and distance between any two anchors in the algorithm, such as $\mathbf{J}$ in Equation (18). We will define a linear operator to describe this operation uniformly.

**Lemma 1** *For a TDoA-based location system, $(p_i, \widehat{\tau}_i)$, $i = 1, 2, \ldots, n$, if there is $\widehat{p}, \epsilon$ to fit*

$$d_{ij} = \widehat{d_{ij}} + \epsilon,$$

then for any $x \in \mathbb{R}^2$, if $p'_i = p_i + x$ and $\widehat{p}'_i = \widehat{p}_i + x$, $i = 1, \ldots, n$, so there is

$$d'_{ij} = \widehat{d}'_{ij} + \epsilon$$

**Definition 2** A linear operator is defined as $T_n : \mathbb{R}^n \to \mathbb{R}^{C_n^2}$, let

$$(T_n x)_k = x_i - x_j, \quad 1 \leq i < j \leq n,$$

where $k = \frac{(2n-i)(j-1)}{2} + j$ is the $k$th sequence of $(i, j)$ with the order $i < j$. For instance, $T_4$ can be expressed as

$$T_4 = \begin{bmatrix} 1 & -1 & 0 & 0 \\ 1 & 0 & -1 & 0 \\ 1 & 0 & 0 & -1 \\ 0 & 1 & -1 & 0 \\ 0 & 1 & 0 & -1 \\ 0 & 0 & 1 & -1 \end{bmatrix}$$

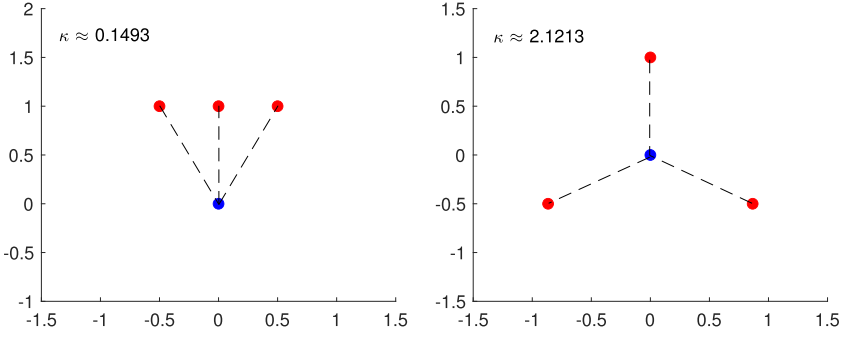

**FIGURE 2** Value in different arrangements of coordinates $\kappa$

with Definition 2, so we could obtain the new **J** in Equation (18) as $\mathbf{J} = T_n[\frac{x-x_i}{d_i} \quad \frac{y-y_i}{d_i}]_{1\leq i \leq n}$.

To estimate the relationship between the algorithm error and the anchor distribution, a new metric is introduced to describe the position of anchors.

**Definition 3** DoP for TDoA. For any set of non-zero coordinates $P = [p_1 \ p_2 \ \ldots \ p_n]^T$, we assume that

$$\kappa(P) = \sigma_{\min}\left(T_n \begin{bmatrix} p_1/\|p_1\| \\ p_2/\|p_2\| \\ \vdots \\ p_n/\|p_n\| \end{bmatrix}\right) \quad (19)$$

where $\sigma_{\min}(\cdot)$ is the minimum singular value of a matrix.

A geometric interpretation of this definition is to describe the angular dispersion for a set of vectors. In general, a higher value indicates a more dispersed angle of the set of vectors. Figure 2 shows an example. The points around the origin are the coordinates of $p_i$s. The more dispersed the coordinates are relative to the origin, the higher the $\kappa$ value is.

Regarding singular values, we have the following properties.

**Proposition 1** For the singular value of the matrix $A \in \mathbb{R}^m \times \mathbb{R}^n$, we have the following properties [29]:

1. $\sigma_{\min}(A) = \min_{\|x\|=1}\|Ax\|$.
2. $\sigma_{\max}(A) = \max_{\|x\|=1}\|Ax\| = \|A\|$.

Based on the properties of singular values, we have the following lemma:

**Lemma 2** Assume that $A, \Delta A \in \mathbb{R}^m \times \mathbb{R}^n$; then,

$$|\sigma_{\min}(A + \Delta A) - \sigma_{\min}(A)| \leq \sigma_{\max}(\Delta A) = \|\Delta A\|$$

## 3.2 | Linear estimator

We suppose there are $n+1$ anchors labelled from $p_0$ to $p_n$. Then by Lemma 1, $p_0$ could be set to the origin. By dividing $r_i r_j$ on both sides of Equation (10), we have

$$\frac{\widehat{d}_{j0}x_i - \widehat{d}_{i0}x_j}{r_i r_j}x + \frac{\widehat{d}_{j0}y_i - \widehat{d}_{i0}y_j}{r_i r_j}y$$
$$= \frac{1}{2r_i r_j}\left(\widehat{d}_{j0}r_i^2 - \widehat{d}_{i0}r_j^2 - \widehat{d}_{i0}\widehat{d}_{j0}\widehat{d}_{ij}\right), \quad \forall \ 1 \leq i < j \leq n. \quad (20)$$

We again call the linear system $Mp = f$. For that system, the condition number of $M$ plays an important role in magnification of the relative error and thus can be considered the DoP, that is,

$$\text{Error of } P = \text{cond } M \cdot \text{Error of } f.$$

For the condition number of $M$, we have the following result:

**Theorem 1** Let $P = [p_1, p_2, \ldots, p_n]^T$; if the tag's location $p = p_0 + \delta$ is located in a neighbourhood near the central anchor $p_0$, then

$$|\sigma_{\min}(M) - \kappa(P)|$$
$$\leq CC'(\delta)\max_i \frac{\|\delta\|}{r_i} + C\max_i \frac{\epsilon_i}{r_i}, \quad C = \sqrt{2}nC_n^2 \quad (21)$$

where $\epsilon_i \sim N(0, c^2\sigma_i^2)$, $\sigma_i$ is the error of the $\widehat{d}_i$, and $C'$ is a constant depending on $\delta$ only.

*Proof:* We first prove the case when $\delta = 0$, that is, the target is also located in the origin. Since $\widehat{d}_{i0} = d_{i0} + \epsilon_i = d_i + \epsilon_i = r_i + \epsilon_i$, $A_{ij}$ and $B_{ij}$ become

$$A_{ij} = \left(\frac{x_i}{r_i} - \frac{x_j}{r_j}\right) + \frac{\epsilon_i \epsilon_j}{r_i r_j}\left(\frac{x_i}{\epsilon_i} - \frac{x_j}{\epsilon_j}\right)$$
$$B_{ij} = \left(\frac{y_i}{r_i} - \frac{y_j}{r_j}\right) + \frac{\epsilon_i \epsilon_j}{r_i r_j}\left(\frac{y_i}{\epsilon_i} - \frac{y_j}{\epsilon_j}\right)$$

By Definition 2, we have

$$M = \begin{bmatrix} A_{ij} & B_{ij} \end{bmatrix}_{1 \le i < j \le n}$$
$$= T_n \text{diag}(\frac{1}{r_i})P + \text{diag}(\frac{\epsilon_i \epsilon_j}{r_i r_j})T_n \text{diag}(\frac{1}{\epsilon_i})P.$$

By Lemma 2, we have

$$\left| \sigma_{\min}(M) - \sigma_{\min}(T_n \text{diag}(\frac{1}{r_i})P) \right| \le \left\| \text{diag}(\frac{\epsilon_i \epsilon_j}{r_i r_j})T_n \text{diag}(\frac{1}{\epsilon_i})P \right\|.$$

Note that $\sigma_{\min}(T_n \text{diag}(\frac{1}{r_i})P) = \kappa(P)$ and

$$\left\| \text{diag}(\frac{\epsilon_i \epsilon_j}{r_i r_j})T_n \text{diag}(\frac{1}{\epsilon_i})P \right\|$$
$$\le \left\| \text{diag}(\frac{\epsilon_i \epsilon_j}{r_i r_j})T_n \text{diag}(\frac{r_i}{\epsilon_i}) \right\| \left\| \text{diag}(\frac{1}{r_i})P \right\|,$$

where $\left\| \text{diag}(\frac{1}{r_i})P \right\| \le n$. On the other hand, for any $x_i^2 + x_j^2 \le 1$, we have

$$\frac{\epsilon_i \epsilon_j}{r_i r_j}\left( \frac{r_i x_i}{\epsilon_i} - \frac{r_j x_j}{\epsilon_j} \right)$$
$$\le \sqrt{2}\, \frac{\epsilon_i \epsilon_j}{r_i r_j} \sqrt{\left(\frac{r_i x_i}{\epsilon_i}\right)^2 + \left(\frac{r_j x_j}{\epsilon_j}\right)^2}$$
$$\le \sqrt{2}\max\left\{ \frac{\epsilon_i}{r_i}, \frac{\epsilon_j}{r_j} \right\} \sqrt{x_i^2 + x_j^2}$$
$$\le \sqrt{2}\max_k \frac{\epsilon_k}{r_k}.$$

Therefore, $\left\| \text{diag}(\frac{\epsilon_i \epsilon_j}{r_i r_j})T_n \text{diag}(\frac{r_i}{\epsilon_i}) \right\| \le \sqrt{2} C_n^2 \max_k \frac{\epsilon_k}{r_k}$. From above, we have

$$|\sigma_{\min}(M) - \kappa(P)| \le C\, \max_k \frac{\epsilon_k}{r_k}, \qquad C = \sqrt{2}n C_n^2. \quad (22)$$

If $\delta > 0$, then $p = \delta$ is located near the neighbour, then by Taylor expansion, we have

$$\begin{aligned} \widehat{d}_{i0} &= d_{i0} + \epsilon_i \\ &= d_i - d_0 + \epsilon = \|p_i - \delta\| - \|\delta\| + \epsilon_i \\ &= r_i + Dd_{i0}(0)\delta + \zeta^T D^2 d_{i0}(0)\zeta + \epsilon_i. \end{aligned} \quad (23)$$

Note that Equation (22) satisfies when $\widehat{d}_{i0} = r_i + \epsilon_i$. Comparing this with Equation (23), if we record $\epsilon_i := Dd_{i0}(0)\delta + \zeta^T D^2 d_{i0}(0)\zeta + \epsilon_i$ in Equation (22), we have

$$|\sigma_{\min}(M) - \kappa(P)| \le C\max_i \frac{Dd_{i0}(0)\delta + \zeta^T D^2 d_{i0}(0)\zeta + \epsilon_i}{r_i}$$
$$\le C\max_i \frac{\|Dd_{i0}(0)\|\|\delta\| + \|D^2 d_{i0}(0)\|\|\delta\|^2 + \epsilon_i}{r_i}$$
$$\le CC'\max_i \frac{\|\delta\|}{r_i} + C\max_i \frac{\epsilon_i}{r_i},$$

where

$$C' = C'(\delta), \ C = \sqrt{2}n C_n^2. \qquad \square$$

**Corollary 1** *Under the condition of* Theorem 1, *the following is true:*

$$\text{cond}(M) = \Theta(\frac{1}{\kappa(P)}).$$

*Proof:* By Theorem 1, we have

$$\kappa(P) + c_0 \ge \sigma_{\min}(M) \ge \kappa(P) - c_0, \quad c_0 \ll \kappa(P).$$

On the other hand, by the proof of Theorem 1,

$$1 \le \sigma_{\max}(M) = \|M\| \le C.$$

Therefore,

$$\text{cond}(M)$$
$$\|$$
$$C'\frac{1}{\kappa(P)} \le \frac{\sigma_{\max}(M)}{\sigma_{\min}(M)} \le C'\frac{1}{\kappa(P)}.$$

This corollary shows that the condition number of the linear estimator is the same order with $\kappa(P)$. Notice the geometric meaning of Definition 3, which means, at least near the base anchor $p_0$, that the DoP of the linear system and the divergence of other anchor positions are directly related. The location of the rest of the anchors is more dispersed, and the system will be better. Note that the role of the central anchor cannot be ignored.

### 3.2.1 | Algorithm without central anchor

The matrix $M$ mentioned in Algorithm 2 has more equations than the system when selecting *any* of the anchors as the centre. Therefore, the condition number of our new $M$ will not be larger than any system with the central anchor.

## 3.3 | Non-linear estimator

Recall that the mathematical model for solving the non-linear estimator is as follows:

$$\underset{p}{argmin}\frac{1}{2}\|r\|^2, \quad r = [d_{ij} - \widehat{d}_{ij}]_{ij}$$

When the iterative algorithm is adopted, the Jacobi matrix **J** of the residual term $r$ has an important effect on the error and stability of the algorithm, namely

$$\text{Error of } p = \sigma^{-1}(\mathbf{J}) \cdot \text{Error of } f.$$

In general, we want the minimum eigenvalue of **J** to have a lower bound, thus ensuring that it is not rank-deficient. Note the form $\mathbf{J}(p)$:

$$\mathbf{J}(p) = \left[\frac{x-x_i}{d_i} - \frac{x-x_j}{d_j} \frac{y-y_i}{d_i} - \frac{y-y_j}{d_j}\right]_{1 \leq i < j \leq n}$$

so we could obtain

$$\sigma_{\min}(\mathbf{J}(p)) = k(P(p))$$

where $P(p) = [p-p_1, p-p_2, \ldots, p-p_n]^T$. Note that $\kappa(P(p))$ is the degree of dispersion with the centre $p$ and the co-ordinates $p_1, p_2, \ldots, p_n$.

## 3.4 | Numerical simulation

When $n = 4$, the anchor deployment is as shown in Figure 3. The left deployment scheme is the one recommended for the linear estimator, and the one to the right respects the non-linear estimator based on the relationship of DoP with $\kappa$. It is conducive to obtain high accuracy of the tag's position with a similar deployment.

It is clear to see that when the rest of the anchors relative to the base anchor are more sub-scattered, a bigger $\kappa$ leads to a lower conditional number. The numerical simulation of the linear algorithm is shown in Figure 4. The solid dots are the locations of the anchors, and the horizontal and vertical co-ordinates, respectively, represent the $x$ and $y$ coordinates of the tag. The one to the left is the recommended deployment, and the one to the right is not a proper deployment, where the condition number is unbounded even inside the rectangular area.

The size of $\kappa$ is directly related to the location error during the positioning process when the non-linear algorithms are chosen, so we demonstrate the size of $\kappa$ when the target is located at a specific position in Figure 5. If the target point and initial guess of the iteration are guaranteed to be located in the region surrounded by the anchors, the solution error can be controlled.

## 4 | EXPERIMENTS

In this section, we conduct experiments for an in-door location system based on UWB signals. The experiments will include both anchor deployments for the linear and non-linear esti-mators and will involve both static-target locating and moving-target tracking problems.

### 4.1 | Experimental settings

There are a total of six positions, and the location of each anchor is shown in Table 1.

For the static-locating problem, the real position of the anchor is shown in Table 2.

For the target tracking problem, we choose the following track for each development, as shown in Table 3.

For the static location, we use the root mean squared error (RMSE) of the Euclidean distance between the actual and estimated position. That is,

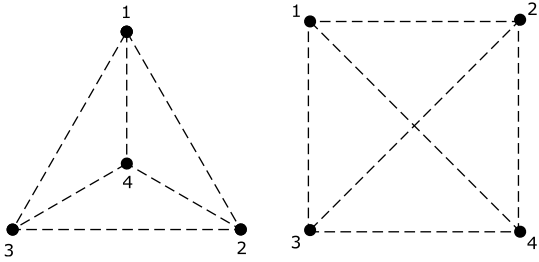

**FIGURE 3** Deployment for linear and non-linear algorithms

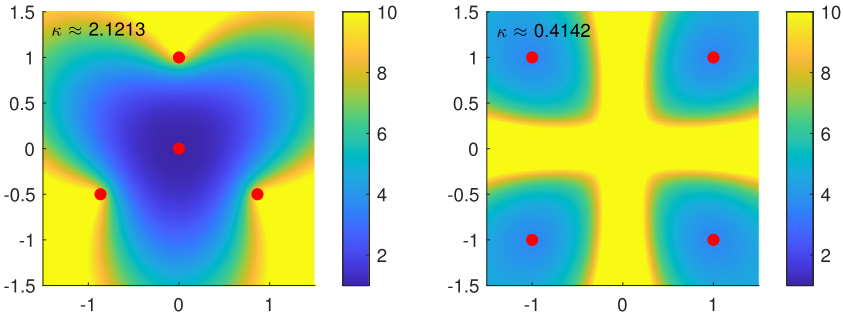

**FIGURE 4** Numerical simulation of linear algorithm

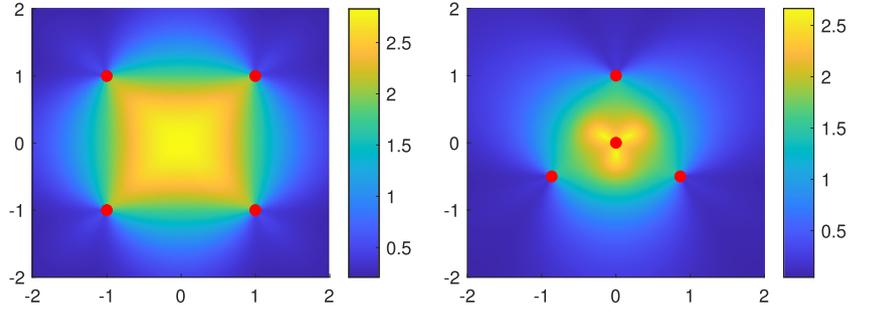

**FIGURE 5** Numerical simulation of non-linear algorithm

**TABLE 1** Positions of anchors

| Anchor label | x position (metre) | y position (metre) |
| --- | --- | --- |
| 1 | 20.961 | 68.941 |
| 2 | 20.911 | 63.929 |
| 3 | 22.652 | 66.441 |
| 4 | 25.417 | 66.461 |
| 5 | 28.274 | 63.910 |
| 6 | 28.324 | 68.961 |

**TABLE 2** Actual positions of targets

| Deployment type | x position (metre) | y position (metre) |
| --- | --- | --- |
| Rectangular deployment | 24.715 | 66.554 |
| Triangular deployment | 22.650 | 66.667 |

**TABLE 3** Actual movements of tracking object moving along straight line (estimated)

| Deployment type | Track segment (from → to) |
| --- | --- |
| Rectangular deployment | (22.11, 68) → (22.11, 64.75) |
| Triangular deployment | (22.03, 64.2) → (22.03, 68.4) |

$$RMSE = \sqrt{\frac{1}{N}\sum_{k=0}^{N}(x_k - x)^2 + (y_k - y)^2},$$

where $N$ is the number of sample points, $(x_k, y_k)$ is the estimated position, and $(x, y)$ is the actual target location.

For the tracking problem, we calculate the RMSE between the estimated location and the distance to the actual moving line segment, namely,

$$RMSE = \sqrt{\frac{1}{N}\sum_{k=0}^{N}d(p_k, L)^2},$$

where $N$ is the number of tracking points, and $d(p_k, L)$ is the Euclidean distance between the estimated point and the line segment.

**TABLE 4** Experimental results for static-target locating problems

| Deployment | Linear estimator | Non-linear estimator |
| --- | --- | --- |
| Rectangular (RMSE) | 0.672 | 0.142 |
| Triangular (RMSE) | 0.127 | 0.114 |
| Improvement (m) | 0.545 | 0.028 |

Abbreviation: RMSE, root mean squared error.

## 4.2 | Location of static targets

In this section, we compare the location error of linear estimator and non-linear estimator using the two deployments proposed in Figure 3 (i.e. triangular and rectangular).

### 4.2.1 | Experimental results

The results for static-target tracking are shown in Table 4. Figure 6 shows the location error of the linear estimator, adopting Algorithm 2. Figure 7 shows the result when adopting the non-linear Algorithm 3.

The experiment has shown that for the linear estimator, when the triangular deployment is adopted, the RMSE is approximately 0.127 m, improved by ∼0.55 m compared with the rectangular deployment with 0.672 m RMSE. This result is consistent with the result in Section 3 and the numerical simulation in Figure 4. This means that we should prefer the deployment with a centering anchor. For the non-linear estimator, both deployments attain a close location error, with only a ∼0.028 m difference.

Comparing the linear and non-linear estimators, the experiment shows that for the triangular deployment, the linear and non-linear location errors are nearly the same. On the other hand, for the rectangular deployment, the linear estimator shows an inferior result. In addition, we could conclude that the non-linear estimator has higher robustness in terms of anchor deployment. However, in practice, the rectangular deployment might serve a larger area compared with that of the triangular deployment.

## 4.3 | Target tracking with KF

In this section, we shall conduct experiments on the tracking of moving targets, which appears frequently in real-time location systems [30]. Under the LoS condition, the received time-stamps for the TDoA measurement constitute a temporal sequence with Gaussian noises, and thus it is suitable to use KF to eliminate the measurement error [31].

### 4.3.1 | Kalman filter

The process for KF is shown below (the input control is omitted).
Prediction phase:

$$\widehat{x}_k = A x_{k-1}$$
$$\widehat{P}_k = A P_{k-1} A^T + Q_k.$$

Correction phase:

$$K_k = \widehat{P}_k H_k^T \left( H_k \widehat{P}_k H_k^T + R_k \right)^{-1}$$
$$x_k = \widehat{x}_k + K_k (z_k - h(x_k))$$
$$P_k = (I - K_k H_k) \widehat{P}_k.$$

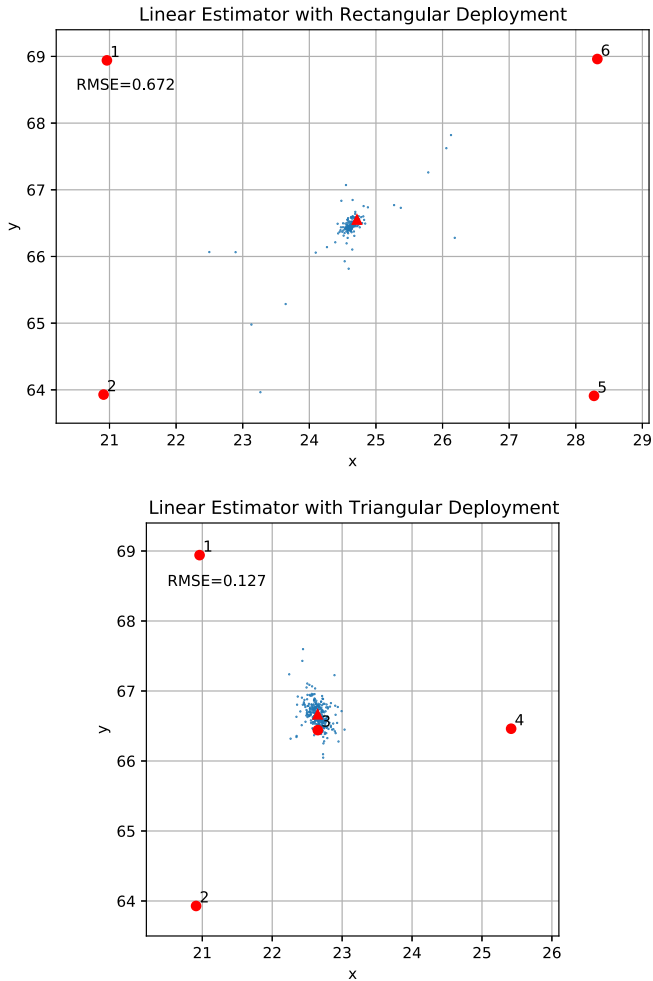

**FIGURE 6** Static location using linear estimator. RMSE, root mean squared error

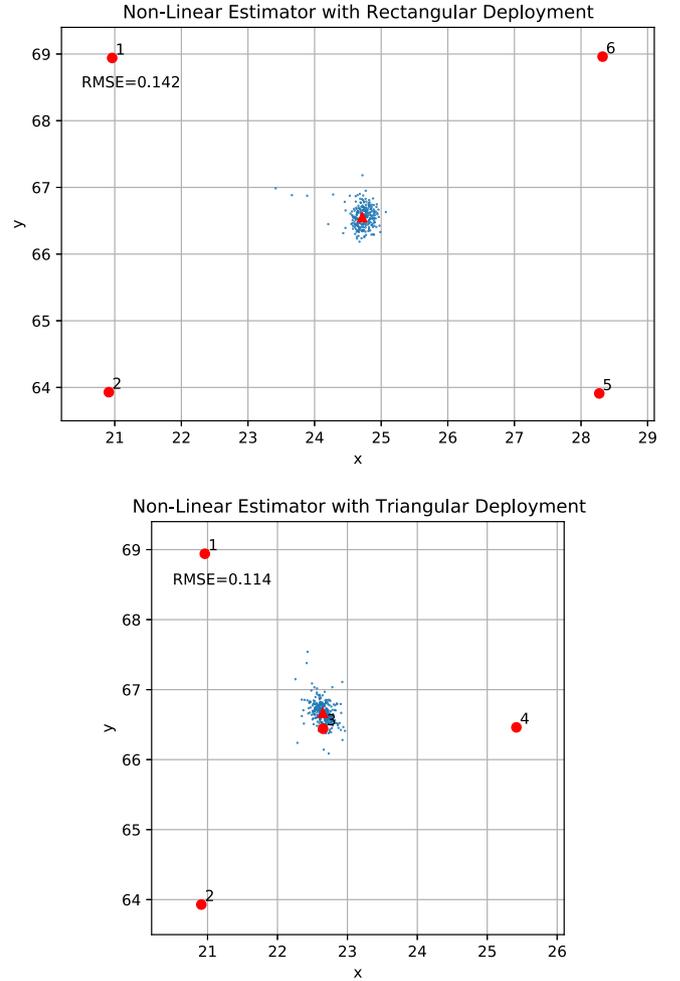

**FIGURE 7** Static location using non-linear estimator. RMSE, root mean squared error

### 4.3.2 | System model

For simplicity, we shall adopt the Brownian model, in which we let $A$ be the identity matrix and the process noise $Q_k = qq^T$, where $q$ approximates the expected movements within the step for each spatial dimension. $R_k$ is the covariance matrix for the measurement. We usually take it as $R_k = r^2 \mathbf{I}$, where $r^2$ is the variance of the measurement, which usually depends on the hardware.

For the linear estimator (Equation (13)), we have the measurement function $h(x) = Mx, H = M$, and measurement $z_k = f_k$.

For the non-linear estimator (Equation (3)), we have the measurement function $h(x) = d_{ij}(x)$. Since the measurement is not a linear function of $x$, the EKF is adopted. In this case, we need to set $H = \mathbf{J}$, which is the Jacobi matrix of $d_{ij}(x)$.

### 4.3.3 | Experimental results

We performed four tracking experiments using the algorithms with different anchor deployments[1]. The results are shown in Figures 8 and 9 and Table 5.

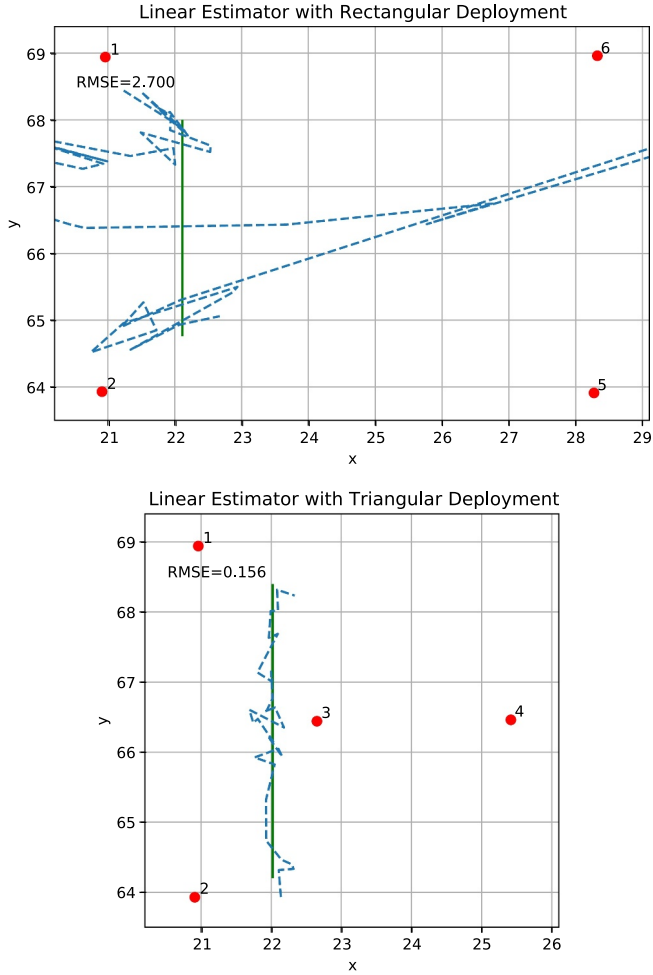

**FIGURE 8** Target tracking using linear estimator. RMSE, root mean squared error

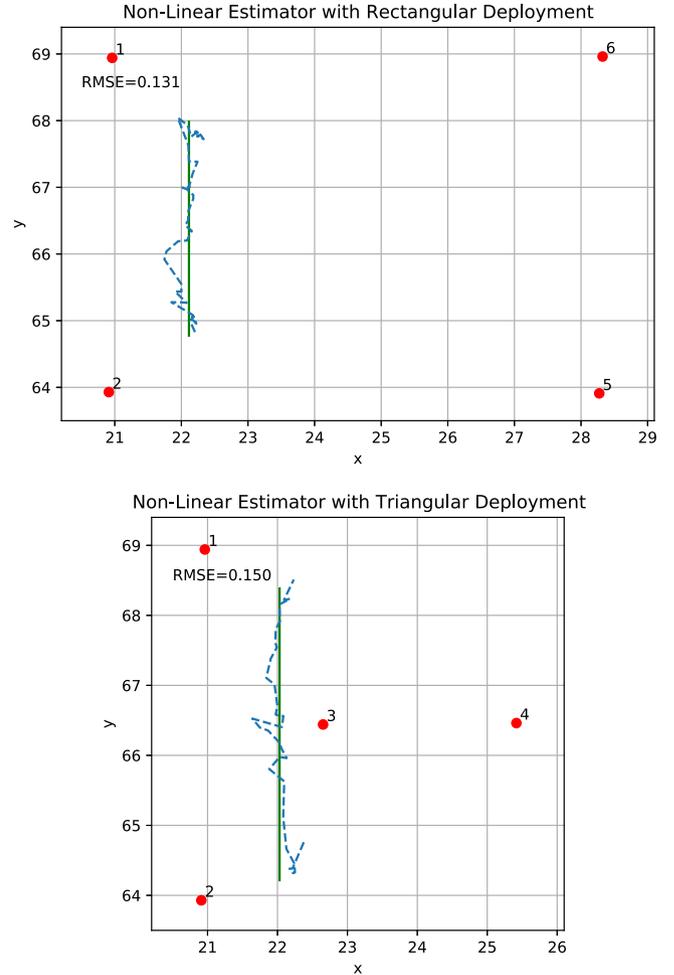

**FIGURE 9** Target tracking using non-linear estimator. RMSE, root mean squared error

It is clear that for the linear estimator, the rectangular deployment attains only ∼2.7 m average accuracy, which implies failure in tracking the target, as illustrated in Figure 4. The triangular deployment reduces the RMSE by ∼2.54 m, whereas the non-linear estimator successfully tracks the target with smaller error (∼0.019 m absolutely difference) with both types of deployment, as desired.

## 5 | CONCLUSION

Two typical algorithms for TDoA location systems are studied, and DoP is analyzed with respect to anchor deployment. Definition of $\kappa$ has a deep relationship with the value of DoP. The geometric meaning of $\kappa$ is the angular dispersion of the anchors, which reveals an intuitive approach to anchor deployment. For the linear estimator, better position estimation can be obtained when a base anchor is given and other anchors are distributed around it at a uniform angle. For the non-linear estimator, if we want to obtain better position estimation results, the anchors need to be distributed around the tag at a uniform angle.

**TABLE 5** Experimental results for target tracking problems

| Deployment | Linear estimator | Non-linear estimator |
| --- | --- | --- |
| Rectangular (RMSE) | 2.700 | 0.131 |
| Triangular (RMSE) | 0.156 | 0.150 |
| Improvement (m) | 2.544 | −0.019 |

Abbreviation: RMSE, root mean squared error.

For static location and tracking problems, the results are consistent with mathematical analyses and numerical simulations. The linear estimator shows sensitivity to anchor arrangements, with a high location error for the rectangular deployment. On the other hand, the non-linear estimator is robust to anchor deployment. Moreover, the rectangular deployment is recommended because it can serve a larger area with the same number of anchors.

For both estimators, anchors are deployed in quite different ways. In practice, if the linear and non-linear algorithms work together, the number of anchors should be increased to ensure that the deployment can meet the requirements of both algorithms simultaneously.

## ACKNOWLEDGMENTS
This research is supported by the National Natural Science Foundation of China under Grant Nos. 61873119 and 61911530247, the Science and Technology Innovation Commission of Shenzhen under Grant Nos. KQJSCX20180 322151418232, and the Scientific Research Foundation of Science and Technology on Near-Surface Detection Laboratory of China under Grant Nos. TCGZ2018A006.

## ORCID
*Fengyun Zhang*  https://orcid.org/0000-0001-8274-0616
*Shuang-Hua Yang*  https://orcid.org/0000-0003-0717-5009

## ENDNOTE
[1] Here we choose the sensor variance $R_k = 0.5\mathbf{I}\ m^2$, and the process variance $q_k = 0.1\ m^2$.